\documentclass[12pt,aps,prc,showpacs,amsmath,showkeys]{revtex4}

\usepackage[dvips]{graphicx}

\usepackage{amssymb}
\usepackage{epstopdf}

\newcommand{\ba}{\begin{eqnarray}}
\newcommand{\ea}{\end{eqnarray}}
\newcommand{\bmath}{\begin{mathletters}}
\newcommand{\emath}{\end{mathletters}}
\newcommand{\ban}{\begin{eqnarray*}}
\newcommand{\ean}{\end{eqnarray*}}

\begin{document}

\title{Relativistic Harmonic Oscillator with Spin Symmetry}
\author{Joseph N. Ginocchio}

   \affiliation{MS B283, Los Alamos National Laboratory,
   Los Alamos, NM, 87545, U.S.A.}

\date{\today}

\begin{abstract}
The eigenfunctions and eigenenergies for a Dirac Hamiltonian with 
equal scalar and vector
harmonic oscillator potentials are derived. Equal scalar and vector 
potentials may be applicable to the
spectrum of an antinucleion imbedded in a nucleus.  Triaxial, axially 
deformed, and spherical oscillator
potentials are considered. The spectrum has a spin symmetry for all 
cases and, for the spherical harmonic
oscillator potential, a higher symmetry analogous to the SU(3) 
symmetry of the non-relativistic harmonic
oscillator is discussed.
\end{abstract}
\maketitle
\section{Introduction}

Recent theoretical investigations have suggested the possibility that 
the lifetime of an
antinucleon inside of a nucleus could be significantly enhanced \cite 
{thomas}. The
relativistic mean fields of antinucleons calculated in a self 
consistent Hartree
approximation of a nuclear field theory \cite {dirk,rein} indicate 
that the scalar
$V_S(\vec r)$ and vector potentials $V_V(\vec r)$ are approximately 
equal. This implies
that the antinucleon spectrum will have an approximate spin symmetry \cite
{tassie,bell}. This conclusion is consistent with the fact that the 
nucleon spectrum has
an approximate pseudospin symmetry \cite  {gino,ami} because the scalar and
vector mean field potentials of a nucleon are approximately equal but 
opposite in sign
and the vector potential changes sign under charge conjugation
\cite {gino1}. In fact, the negative energy states of the nucleon do 
show a strong spin
symmetry \cite {meng}. Of course, whether such states could be 
observable can only be
reliably estimated if the antinucleon annihilation potential is 
included in the mean
field calculations.

In this paper we shall solve for the eigenfunctions
and eigenenergies of the triaxial, axially deformed, and spherical 
relativistic harmonic
oscillator for equal scalar and vector potentials with the 
expectation that these
results could be helpful in drawing conclusions about the feasibility 
of observing the
spectrum of an antinucleon in a nuclear environment. The spherical relativistic 
harmonic oscillator with spin symmetry \cite {rav,bahdri,tegen,cent,kukulin,chao} and
pseudospin symmetry \cite {chen,manuel} have been studied previously,
but in this paper we derive the eigenfunctions and eigenenergies for the
triaxial and axially deformed harmonic oscillator as well.

\section{The Dirac Hamiltonian and Spin Symmetry}

The Dirac Hamiltonian, $H$, with an external scalar, $V_S(\vec
r)$, and vector, $V_V(\vec r)$, potential is given by:
\ba
H =\mbox{\boldmath $\alpha\cdot p$}
+ \beta (M + V_S(\vec r)) + V_V(\vec r))  ~,
\label {dirac}
\ea
where \mbox{\boldmath $\alpha$}, $\beta $ are the usual Dirac
matrices, $M$ is the nucleon mass and we set $c$ = 1. The Dirac Hamiltonian is
invariant under a $SU(2)$ algebra for  two limits: $V_S(\vec r) =
V_V(\vec r)  + C_s$ and  $V_S(\vec r) = - V_V(\vec r)  + C_{ps}$
where $C_s,C_{ps}$ are constants \cite{bell}. The former limit has
application to the spectrum of mesons for  which the spin-orbit
splitting is small \cite{page} and for the spectrum of an
antinucleon in the mean field of nucleons \cite{gino1,meng}. The
latter limit leads to pseudospin symmetry in nuclei \cite{gino}.
This symmetry occurs independent of the shape of the nucleus:
spherical, axial deformed, or triaxial.

\subsection{Spin Symmetry Generators}

The generators for the spin $SU(2)$ algebra,
${{S}}_q$, which commute with the Dirac Hamiltonian,
$[\,H_{s}\,,\, {{S}}_q\,] = 0$, for the spin
symmetry limit $V_S(\vec r)= V_V(\vec r)+ C_{s}$ , are given by
\cite{bell}
\ba
{{ {S}}}_q =
\left (
\begin{array}{cc}
 {s}_q &  0 \\
0 & { {\tilde s}}_q
\end{array}
\right )
= \left (
\begin{array}{cc}
  { s}_q  & 0 \\
0 & U_p\,{ s}_q \, U_p
\end{array}
\right )
\label{psg}
\ea
where ${ s}_q = \sigma_q/2$ are the usual spin generators,
$\sigma_q$ the Pauli matrices, and $U_p = \, {\mbox{\boldmath
$\sigma\cdot p$} \over p}$ is the momentum-helicity unitary
operator \cite {draayer}. Thus the operators
${ S}_i$ generate an SU(2) invariant symmetry of $H_{s}$.
Therefore each eigenstate of the Dirac Hamiltonian has a partner
with the same energy,
\begin{equation}
H_{s}\ \Phi_{k,{ \mu}}^{s}({\vec r}) = E_k \Phi_{k,{
\mu}}^{s}({\vec r})
\label{eigen}
\end{equation}
where $k$ are the other quantum numbers and ${ \mu} = \pm
{1\over 2}$ is the eigenvalue of
${ S}_z$,
\begin{equation}
{  S}_z\ \Phi_{k,{\mu}}^{s}({\vec r}) = { \mu} \
\Phi_{k,{ \mu}}^{s}({\vec
r}).
\label{Sz}
\end{equation}
The eigenstates in the doublet will be connected by the generators
${ S}_{\pm 1}$,
\begin{equation}
{ S}_{\pm }\  \Phi_{k,{ \mu}}^{s}({\vec
r}) =
\sqrt{{\left ({1 \over 2}
\mp {\tilde \mu} \right )  ({3 \over 2} \pm {\tilde \mu} )}}  \
\Phi_{k,{ \mu} \pm 1}^{s}({\vec r}).
\label{S+}
\end{equation}
The fact that Dirac eigenfunctions belong to the spinor
representation of the spin SU(2), as given in Eqs. (\ref {Sz}) - (\ref{S+}), 
leads to conditions
on the Dirac amplitudes \cite  {mad,ami1,gino2,witek}.

\subsection{Dirac Eigenfunctions and Spin Symmetry}

The Dirac egenfunction can be written as a four dimensional vector
\begin{equation}
\Phi_{k,\mu}({\vec r}) = \left( \begin{array}{c}
g^+_{k,\mu}({\vec r}) \\
g^-_{k,\mu}({\vec r})\\ if^+_{k,\mu}({\vec r})\\if^-_{k,\mu}({\vec 
r}) \end{array}
    \right),
\label {four}
\end{equation}
where $g^{\pm}_{k,\mu}({\vec r})$ are the ``upper Dirac components"
where $+$ indicates spin up and $-$ spin down and
$f^{\pm}_{k,\mu}({\vec r})$ are the ``lower Dirac components" where
$+$ indicates spin up and $-$ spin down. However spin symmetry 
imposes conditions on
these eigenfunctions \cite {gino2,witek} which are derived from 
Equations (\ref {Sz}) -
(\ref {S+})
\bmath
\ba
g^+_{k,{1\over2}}({\vec r}) = g^-_{k,-{1\over2}}({\vec r}) = 
g_{k}({\vec r}), \\
g^+_{k,-{1\over2}}({\vec r}) = g^-_{k,{1\over2}}({\vec r}) = 
0, \\
{f}^{+}_{k,{ 1\over 2}}({\vec r}) = -{f}^{-}_{k,- {
1\over 2}}({\vec r}) = {f}_{k}({\vec r}),\\
({\partial\over \partial x} + i{\partial\over \partial x})\ 
{f}^{+}_{k,- { 1\over
2}}({\vec r}) = ({\partial\over \partial x} - i{\partial\over 
\partial x})\ f^{-}_{k, {
1\over 2}}({\vec r}),\\ \ {\partial\over \partial z}\ {f}^{\pm}_{k,\mp { 1\over
2}}({\vec r}) =\pm ({\partial\over \partial x} \mp i{\partial\over \partial
x})\ f^{\pm}_{k,\pm{ 1\over 2}}({\vec r}).
\ea
\label{cond}
\emath
Thus for spin symmetry the Dirac spin doublets are
\bmath
\ba
\Phi^s_{k,{ 1\over 2}}({\vec r}) = \left( \begin{array}{c}
g_{k}({\vec r}) \\
0\\ if_{k}({\vec r})\\if^-_{k,{ 1\over 2}}({\vec r}) \end{array}
    \right),\
\Phi^s_{k,-{ 1\over 2}}({\vec r}) = \left( \begin{array}{c}
0 \\
g_{k}({\vec r})\\if^+_{k,-{ 1\over 2}}({\vec r}) \\-if_{k}({\vec r})) 
\end{array}
    \right).
\ea
\label{spinef}
\emath
\subsection{Second Order Differential Equation for the Eigenfunctions}
The Dirac Hamiltonian, (\ref {dirac}), gives first order differential 
relations between
${ g}_{k}({\vec r})$ and ${ f}_{k}({\vec r})$ and ${ f}^{\pm}_{k, \mp 
{1\over 2}}({\vec
r})$. In the usual way we turn these equations into a second order 
differential relation
for the upper component. In the limit of spin symmetry this second 
order equation
becomes
\begin{equation}
\left [ p^2 + 2\ ({\tilde E}_k + {\tilde M})\ V(\vec r) - {\tilde E}_k^2 
+ {\tilde M}^2
\right ]{g}_{k}({\vec r}) = 0
\label {second}
\end{equation}
where $V_S(\vec r) = {V(\vec r) } +
V_S^0$, $V_V(\vec r) = {V(\vec r) } +
V_V^0$, ${\tilde M} = M + V_S^0$, and ${\tilde E}_k = E_k - V_V^0$. From (\ref
{dirac}) and setting $\hbar$= 1, the lower components become
\bmath
\ba
f_k(\vec r) = {-1 \over ({\tilde M} + {\tilde E})}\ {\partial \over 
\partial z}\
g_k(\vec r),\\ f_{k,\mp {1 \over 2}}^{\pm}(\vec r) = {-1 \over 
({\tilde M} + {\tilde
E})}\ ({\partial \over \partial x} \mp i\ {\partial
\over
\partial y} )\ g_k(\vec r).
\ea
\label{spinf}
\emath
These relations are consistent with the conditions on the 
eigenfunctions imposed by
spin symmetry in Equations (\ref {cond}).

Equation (\ref {second}) is
basically the energy-dependent Schr\"odinger equation without any 
spin dependence. Hence any potential,
which can be solved analytically with the non-relativistic 
Schroedinger equation, is solvable in the spin
limit, but the energy spectrum will be different because of the 
nonlinear dependence on the energy.
Hence the Coulomb and harmonic oscillator are solvable. The Coulomb 
potential has been
solved (for general scalar and vector potentials) \cite {gino} and applied to 
meson spectroscopy
\cite {page}. More appropriate for nuclei is the harmonic oscillator.

\section{Harmonic Oscillator}

First we shall discuss the triaxial harmonic oscillator, then the 
axially symmetric
harmonic oscillator, and finally the spherical harmonic oscillator.

\subsection{Triaxial Harmonic Oscillator}

For the potential $V(\vec r) = {{\tilde M} \over 2}(\omega_1^2 \ x^2 
+ \omega_2^2\  y^2
+\omega_3^2 \ z^2) $, the second order differential equation (\ref 
{second}) becomes
\begin{equation}
\left [ {{\partial}^2\over {\partial x^2}} + {{\partial}^2\over 
{\partial y^2}} +
{{\partial}^2\over {\partial z^2}} - {({\tilde E}_{n_1,n_2,n_3} + 
{\tilde M})\ {\tilde
M}
}(\omega_1^2 \ x^2 + \omega_2^2\  y^2 +\omega_3^2 \ z^2)+ {\tilde
E}_{n_1,n_2,n_3}^2 - {\tilde M}^2
\right ]{g}_{{n_1,n_2,n_3}}({\vec r}) = 0 .
\label {secondho}
\end{equation}
\subsubsection{Eigenfunctions}
Introducing the product ansatz for the eigenfunction
${g}_{{n_1,n_2,n_3}}({\vec r}) \propto {g}_{n_1}({x})\ {g}_{n_2}({y})\ 
{g}_{n_3}({z})$, we
derive the three equations
\begin{equation}
\left [ {{\partial}^2\over {\partial x_i^2}}   -  x_i^2 + 2n_i + 1 \right ]\
{g}_{n_i}({x_i})
  = 0 ,
\label {tho}
\end{equation}
where $x_1 = \lambda_1\ x,\  x_2 = \lambda_2\ y,\  x_3 = \lambda_3\ z, $ and
\bmath
\ba
\lambda_i = \left [ {({\tilde E}_{n_1,n_2,n_3} + {\tilde M})\ {\tilde 
M}\ \omega_i^2
} \right ]^{1\over 4},\\ {\tilde E}_{n_1,n_2,n_3}^2 - {\tilde M}^2 = 2\
\sum_i\lambda_i^2\ (n_i + {1\over 2})
\ea
\label{tri}
\emath
The bound eigenstates are given by
\begin{equation}
{g}_{{n_1,n_2,n_3}}({\vec r}) = {\cal N}(E_{n_1,n_2,n_3})\ {g}_{{n_1}}(x_1)\
{g}_{{n_2}}(x_2)\ {g}_{{n_3}}(x_3)  ,
\label {ef}
\end{equation}
where
\begin{equation}
{g}_{{n}}({x}_i) = 
\sqrt{\lambda_i\over
\sqrt{\pi}\ 2^{n_i}\ n_i!}\ e^{-x_i^2 \over 2}\ H_{n_i}(x_i) ,
\label {gi}
\end{equation}
and $H_{n_i}(x_i) $ is the Hermite polynomial which means that
${g}_{{n_1,n_2,n_3}}({\vec r})$ has $n_1,n_2,n_3$ nodes in the
$x, y,$ and $z$ directions, respectively.
${\cal N}(E_{n_1,n_2,n_3})$ is the normalization determined by
$\int_{-\infty}^{\infty}dx
\int_{-\infty}^{\infty}dy\int_{-\infty}^{\infty}dz\ 
(|{g}_{{n_1,n_2,n_3}}({\vec r})|^2
+ |{f}^+_{{n_1,n_2,n_3,\mu}}({\vec r})|^2 + |{f}^-_{{n_1,n_2,n_3,\mu}}({\vec r})|^2) = 1$.

 From (\ref {spinf}) the lower components are

\bmath
\ba
{f}_{{n_1,n_2,n_3}}({\vec r}) = {\cal N}(E_{n_1,n_2,n_3})\
  {g}_{{n_1}}({x_1}) \ {g}_{{n_2}}({x_2})\  f_{n_3 + 1}(x_3),\\
{f}_{{n_1,n_2,n_3},\mp{1\over2}}^{(\pm)}({\vec r}) = {\cal N}(E_{n_1,n_2,n_3})\
({f}_{{n_1}+1}({x_1}) \ {g}_{{n_2}}({x_2}) \mp i\ {g}_{{n_1}}({x_1}) \
{f}_{{n_2}+1}({x_2}))\  g_{n_3}(x_3),
\label {eff}
\ea
\emath
where

\ba
f_{n_i + 1}(x_i)= \lambda_i\ \sqrt{\lambda_i\over
\sqrt{\pi}\ 2^{n_i}\ n_i!}\ e^{-x_i^2}\ ({H_{n_i+1}(x_i) \over 2} - 
n\ H_{n_i-1}(x_i)).
\label {fi}
\ea

Clearly the function $f_{n + 1}(x)$ is a polynomial of order $n + 1$. 
Evaluation for low
$n$ demonstrates that it has $n + 1$ nodes so we assume that it has 
$n + 1$ nodes for
all $n$. This means that ${f}_{{n_1,n_2,n_3}}({\vec r})$ has one more 
node in the
$z$-direction than ${g}_{{n_1,n_2,n_3}}({\vec r})$ and the same 
number of nodes in the $x$
and $y$ directions as ${g}_{{n_1,n_2,n_3}}({\vec r})$. On the other 
hand the amplitudes
${f}_{{n_1,n_2,n_3},\mp{1\over2}}^{(\pm)}({\vec r})$ have the same 
number of nodes in
the z-direction as ${g}_{{n_1,n_2,n_3}}({\vec r})$.

Using these amplitudes the normalization becomes
\begin{equation}
  {\cal N}(E) = \sqrt{{2\ ({\tilde E} + {\tilde M}) \over 3
{\tilde E} + {\tilde M}}}\ .
\label {norm}
\end{equation}

\subsubsection{Eigenenergies}
 From Equations (\ref {tri}) the eigenvalue equation is

\begin{equation}
\sqrt{{\cal E}_{n_1,n_2,n_3} + 1}\ ({\cal E}_{n_1,n_2,n_3} - 1) = 
\sum_{i=1}^3 C_i\
(n_i + {1 \over 2}),
\end{equation}
where ${\cal E}_{n_1,n_2,n_3} = {{\tilde E}_{n_1,n_2,n_3} \over 
{\tilde M}},C_i = 2\ {\omega_i \over {\tilde M}}$, and $n_i = 0, 1,\dots$.
Thus both the eigenfunctions and the eigenenergies are independent of spin.

This
eigenvalue equation is solved on Mathematica:
\begin{equation}
E_{n_1,n_2,n_3} =  {\tilde M } \left [ B(A_{n_1,n_2,n_3}) + {1 \over 3} + 
{4\over 9\ B(A_{n_1,n_2,n_3})}
\right] + V_V^0 ,
\label {E1}
\end{equation}
where
\begin{equation}
B(A_{n_1,n_2,n_3}) =   \left [ A_{n_1,n_2,n_3} + \sqrt{A_{n_1,n_2,n_3}^2
- {32 \over 27}
\over 2} \right
]^{{2\over 3}} ,
\label {B}
\end{equation}
and $A_{n_1,n_2,n_3} = \sum_{i=1}^3 C_i\ (n_i + {1 \over 2})$.

The eigenvalues $E_{n_1,n_2,n_3}$ are  real for all values of 
${n_1,n_2,n_3}$ as
long as  $C_i,V^0_{S,V}$ are real. Although true it is not obvious because
$B(A_{n_1,n_2,n_3})$ is not real for all
$ A_{n_1,n_2,n_3}$ real. From (\ref {B}),
$B(A_{n_1,n_2,n_3})$ is clearly complex for
$A_{n_1,n_2,n_3} <
\sqrt{ 32 \over 27}$. However, we now show analytically that 
$E_{n_1,n_2,n_3}$ will
still be real even if $B(A_{n_1,n_2,n_3})$ is complex as long as $|B(A_{n_1,n_2,n_3})| = 2/3$.

The imaginary part of ${ E}_{n_1,n_2,n_3}$ is
\begin{equation}
Im \ { E}_{n_1,n_2,n_3} = {\left [ B(A_{n_1,n_2,n_3}) + {1\over 3} + 
{4\over 9\ B(A_{n_1,n_2,n_3})}
\right] - \left [ B(A_{n_1,n_2,n_3}) + {1\over 3} + {4\over 9\ B(A_{n_1,n_2,n_3})}
\right]^*\over 2i}
\end{equation}
Writing $B(A_{n_1,n_2,n_3}) = |B(A_{n_1,n_2,n_3})|e^{i\psi}$
\begin{equation}
Im \ { E}_{n_1,n_2,n_3} = \left ( |B(A_{n_1,n_2,n_3})|- {4\over 9\ 
|B(A_{n_1,n_2,n_3})|}\right ) sin \psi,
\end{equation}
and therefore ${ E}_{n_1,n_2,n_3}$ is real if $|B(A_{n_1,n_2,n_3})| = 2/3$ 
independent of $\psi$.
One can show
numerically that $|B(A_{n_1,n_2,n_3})| = 2/3$ for {\it all}
$ A_{n_1,n_2,n_3} $ in the range from zero to $\sqrt{{ 32 \over 27}}$, $0 \le A_{n_1,n_2,n_3} 
\le \sqrt{{ 32
\over 27}}$. For $A_{n_1,n_2,n_3} \ge \sqrt{{ 32 \over 27}}$, $B(A_{n_1,n_2,n_3})$ is clearly 
real and hence
${ E}_{n_1,n_2,n_3}$ is real.

The spectrum is non-linear in contrast to the non-relativistic harmonic
oscillator. However for small $A_{n_1,n_2,n_3}$
\begin{equation}
E_{n_1,n_2,n_3} \approx {\tilde M}\ ( 1 + {A_{n_1,n_2,n_3}\over \sqrt{2}} + 
\cdots) + V^0_V
\label {pert}
\end{equation}
and therefore the binding energy, $E_{n_1,n_2,n_3} - {\tilde M} \approx
\sum_{i=1}^3 \omega_i \ (n_i + {1 \over 2})+ V_V^0$, in agreement with the
non-relativistic harmonic oscillator. For large $A_{n_1,n_2,n_3}$ the spectrum goes as
\begin{equation}
E_{n_1,n_2,n_3} \approx {\tilde M}\ (  A_{n_1,n_2,n_3}^{2\over 3} + {1 \over 3} + \cdots) +
V^0_V,
\label {asymp}
\end{equation}
which, in lowest order, agrees with the spectrum for ${\tilde 
M}\rightarrow$  0 \cite{bahdri}.

\subsection{Axially Symmetric Harmonic Oscillator}

For the axially symmetric harmonic oscillator  $\omega_1 = \omega_2
=\omega_{\perp}$ and, hence, the potential depends only on 
$r_{\perp}$ and not the
azimuthal angle $\phi$,
$V(\vec r) = {{\tilde M}
\over 2}(\omega_{\perp}^2\
r_{\perp}^2 +\omega_3^2 \ z^2) $, where $x =r_{\perp}\ cos\phi,y = 
r_{\perp}\ sin\phi$.
This independence of the potential on $\phi$ implies that the Dirac 
Hamiltionian is invariant under
rotations about the $z$-axis, $[H_s,L_z]=0$, where
\begin{equation}
{ { L_z}} = \left( \begin{array}{cc}
{{\ell_z}} & 0 \\
0 & { {{\tilde \ell_z}}}  \end{array}
    \right),
\label{Lz}
\end{equation}
and ${\vec \ell} = {\vec r} \times {\vec p}$ and and ${{\vec {\tilde 
\ell}}}  = U_p\  {\vec\ell}\ U_p$.
The Dirac eigenstates will then be an
eigenfunction of $L_z$ and $J_z = L_z + S_z$.

\bmath
\ba
L_z  \Phi_{N, n_3, \Lambda, \mu}^{s}({\vec
r}) =
\Lambda  \
\Phi_{N, n_3, \Lambda, \mu}^{s}({\vec r}),\\
J_z\  \Phi_{N, n_3, \Lambda, \mu}({\vec r})= \Omega  \  \Phi_{N, n_3, 
\Lambda, \mu }^{s}({\vec r}),\  \Omega
= \Lambda + \mu,
\label{JJ}
\ea
\emath
where $N$ is the total harmonic oscillator quantum number and $n_3$ is the 
number of harmonic oscillator quanta in the $z$ direction, and are discussed in more
detail below.
\subsubsection{Eigenfunctions}

Since the potential has no $\phi$ dependence the second
order differential equation (\ref  {second}) separates into an 
equation  for $x_3$
and an equation for $\rho$
\bmath
\ba
\left [ {{\partial}^2\over {\partial {x_3}^2}}   -  x_3^2 + 2n_3 + 1 \right ]\
{g}_{n_3}({x_3})
  = 0 ,\\
\left [ {{\partial}^2\over {\partial {\rho}^2}} + {1-4\ \Lambda^2 
\over 4\ \rho^2}   -
\rho^2 + 2\ n_{\perp} + 2
\right ]\ \sqrt{\rho}\ {g}_{n_{\rho}}({\rho})
  = 0 ,\\
\label {as}
\ea
\emath
where $\rho = \lambda_{\perp}\ r_{\perp}, n_{\perp} = 2n_{\rho} + \Lambda$,
and
\bmath
\ba
g_{N,n_3,\Lambda}(\vec r) = g_{n_{\rho}}(\rho)\ g_{n_3}(z)\ 
e^{i\Lambda \phi},\\  \
\lambda_{\perp} =   \left [ {({\tilde E}_{N,n_3} + {\tilde M})\ {\tilde M}\
\omega_{\perp}^2} \right ]^{1\over 4}, \\ {\tilde E}_{N,n_3}^2 - {\tilde M}^2 = 2\
(\lambda_{\perp}^2\ (\ n_{\perp} + 1) + \lambda_{3}^2\ (\ n_3 + {1 \over 2})).
\ea
\label{def}
\emath
The quantum numbers $N,n_{\perp},n_3,\Lambda$ are the ``asymptotic" quantum 
numbers \cite {bohr},
where $N= n_{\perp} + n_3$ is the total number of oscillator quanta, 
$N = 0, 1, \dots $, $n_{\perp}$ is the number
of oscillator quanta in the $r_{\perp}$-direction, $n_{\perp} = 0, 1, \dots N$, $n_3$
is the number of oscillator quanta in the $z$-direction, $n_3 = 0, 1, \dots N$, and 
$\Lambda$ is the angular momentum along
the z-axis, $\Lambda = \pm n_{\perp},  \pm (n_{\perp} - 2) \dots \pm 
1, 0 $. The upper components of the
eigenstates are given by

\ba
{g}_{N,n_3,\Lambda}({\vec r}) = {\cal N}(E_{N,n_3})\ \lambda_{\rho}
\sqrt{\lambda_3\ n_{\rho}!\over
\pi \sqrt{\pi}\ 2^{n_3}\ (n_{\rho} +\Lambda )!\ n_3!}\ e^{-{( x_3^2 
+\rho^2) \over 2}}\
\rho^{{\Lambda }}\ L^{(\Lambda)}_{n_{\rho}}(\rho^2)\ e^{i\Lambda 
\phi}\  H_{n_3}(x_3) ,
\label {efa}
\ea
where $ L^{(\Lambda)}_{n_{\rho}}(\rho^2) $ is the Laguerre polynomial 
which means that
${g}_{N,n_3,\Lambda}({\vec r})$ has $n_{\rho}, n_3$ nodes in the
$r_{\perp}$ and $z$ directions, respectively.
${\cal N}(E_{N,n_3})$ is the normalization determined by $\int_0^{2\pi} d\phi\
\int_0^{\infty} \ r_{\perp}\ dr_{\perp} \
\int_{-\infty}^{\infty} \ dz(|{g}_{N,n_3,\Lambda}({\vec r})|^2 +
|{f}^+_{N,n_3,\Lambda,\mu}({\vec r})|^2 + |{f}^-_{N,n_3,\Lambda,\mu}({\vec r})|^2)$ = 1 and is the same function as 
given in (\ref {norm}). These upper
components do not depend on the orientation of the spin.

The lower components are

\bmath
\ba
{f}_{N,n_3,\Lambda}({\vec r}) & = & -{\ {\cal N}(E_{N,n_3,\Lambda})\over
{\tilde M} + {\tilde E_{N,n_3}}}\
\ \lambda_{\rho}^2\ \sqrt{ n_{\rho}!\over
\pi \ (n_{\rho} +\Lambda )!\ }\ e^{-{\rho^2 \over 2}}\
\rho^{{\Lambda }}\ L^{(\Lambda)}_{n_{\rho}}(\rho^2)\ e^{i\Lambda 
\phi}\  f_{n_3 +
1}(x_3),
\ea
\ba
{f}_{N,n_3,\Lambda,-{1\over 2}}^+({\vec r}) \nonumber
&=&
-{{ {\cal N}(E_{N,n_3}})\over
{\tilde M} + {\tilde E_{N,n_3}}}\
\ \lambda_{\rho}^2\ \sqrt{  n_{\rho}!\over
\pi  (n_{\rho} +\Lambda )!\ }\ e^{-{\rho^2 \over 2}}\
\rho^{{\Lambda }-1}\nonumber \\&&( (n_{\rho} + 1)\ L^{(\Lambda - 
1)}_{n_{\rho}+ 1}(\rho^2)+ (n_{\rho} +
\Lambda)\  L^{(\Lambda - 1)}_{n_{\rho}}(\rho^2))\ e^{i(\Lambda - 1) 
\phi}\  g_{n_3}(x_3),
\ea
\ba
f_{N,n_3,\Lambda,{1\over 2}}^-({\vec r}) &= &{ {\cal N}(E_{N,n_3})\over
{\tilde M} + {\tilde E_{N,n_3}}}\
\ \lambda_{\rho}^2\ \sqrt{ n_{\rho}!\over
\pi \ (n_{\rho} +\Lambda )!\ }\ e^{-{\rho^2 \over 2}}\
\rho^{{\Lambda } + 1}( L^{(\Lambda + 1)}_{n_{\rho} }(\rho^2) \nonumber \\
&&+ L^{(\Lambda + 1)}_{n_{\rho} - 1}(\rho^2))\
e^{i(\Lambda + 1)
\phi}\  g_{n_3}(x_3),
\label {effa}
\ea
\emath
where $g_{n_3}(x_3),f_{n_3}(x_3)$ are defined in (\ref {gi}) and (\ref {fi}).
Clearly the function ${f}_{N,n_3,\Lambda}({\vec r})$ has $n_3 + 1$ nodes in
the $z$-direction, one more node than the upper component, but the 
same number of
nodes in the $r_{\perp}$ direction. The amplitude
${f}_{N,n_3,\Lambda,-{1\over 2}}^+({\vec r}))$ has the same number of 
nodes in the $z$ - direction
as the upper component. However, it has $n_{\rho} + 1$ nodes in the 
$r_{\perp}$-direction for low
$n_{\rho}$ and we assume it has $n_{\rho} + 1$ node for all 
$n_{\rho}$; that is, one more node in the
$r_{\perp}$-direction than the upper component. On the other hand the 
amplitude ${f}_{N,n_3,\Lambda,{1\over
2}}^-({\vec r})$ has the same number of nodes in the $z$-direction and the
$r_{\perp}$-direction as the upper component.

\subsubsection{Eigenenergies}
 From Equations (\ref {def}) the eigenvalue equation is

\begin{equation}
\sqrt{{\cal E}_{N,n_3} + 1}\ ({\cal E}_{N,n_3} - 1) =
C_{\perp}\  (n_{\perp} + 1) + C_3\ (n_3 +{1\over 2}),
\label{axen}
\end{equation}
where ${\cal E}_{N,n_3} = {{\tilde E}_{N,n_3} \over {\tilde 
M}},C_{\perp} ={ {2}\
\omega_{\perp}
\over {\tilde M}}$, and
$n_{\rho},n_3 = 0, 1,\dots$. Thus the eigenenergies not only have a degeneracy
due to spin symmetry but they have an additional degeneracy in that 
they only depend on $N$ and $n_3$ and
not on $\Lambda$.

The discussion about eigenergies is the same as for triaxial nuclei and the energy spectrum is
given by 

\begin{equation}
E_{N,n_3} =  {\tilde M} \left [ B(A_{n_{\perp}, n_3}) + {1\over 3} + {4\over 9\ B(A_{n_{\perp},
n_3})}
\right] + V_V^0 ,
\label {E1ax}
\end{equation}
where $A_{n_{\perp}, n_3} =C_{\perp}\  ( n_{\perp}  + 1) + C_3\ (n_3 +{1\over 2})$.

\subsection{Spherical Harmonic Oscillator}

For a spherical harmonic oscillator $\omega_i = \omega$ and hence the 
potential depends only on the radial
coordinate, $ r= \sqrt{x^2 + y^2 + z^2}$, and is independent of the 
polar angle, $\theta$, $z = r\
cos(\theta)$, as well as $\phi$.  The Dirac Hamiltonian will be 
invariant with respect to
rotations about all three axes,
$[L_i, H_{s}]= 0$ where
\begin{equation}
{ {\vec L}} = \left( \begin{array}{cc}
{{\vec\ell}} & 0 \\
0 & { {\vec{\tilde \ell}}}  \end{array}
    \right),
\label{L}
\end{equation}
and hence invariant with respect to a $SU_L(2) \times SU(2)$ group 
where $SU_L(2)$ is generated by the
orbital angular momentum operators $\vec L$. Since the total angular 
momentum, ${\vec J} = {\vec L} +
{\vec S}$, is also conserved, rather than using the four row basis 
for this eigenfunction, it is more
convenient to introduce the spin function $\chi_{\mu}$ explicitly. The
states that are a degenerate doublet are then the states with $j = \ell \pm
{1\over2}$ and they have the two row form \cite {gino2}:

\begin{equation}
\Psi^{s}_{n_r, \ell,j, M}({\vec r}) = \ \left(
\begin{array}{c}
g_{n_r,\ell}(r) \ [Y^{(\ell)} (\theta, \phi)\ \chi]^{(j)}_M
\\ if_{n_r, \ell, j}(r)\ [Y^{(\ell_j)} (\theta,
\phi)\ \chi]^{(j)}_M
\end{array}
\right),
\label{ssph}
\end{equation}
where $\ell_j = \ell\pm 1$ for $j = \ell\pm {1\over 2}$, $Y^{(\ell)}_{m}(\theta,\phi)$ is the spherical harmonic of 
order $\ell$, $n_r$ is the number of radial
nodes of the upper amplitude, and $[Y^{(\ell)} (\theta,
\phi)\ \chi]^{(j)}_M$ is the coupled amplitude $\sum_{m \mu}C^{\ell 
{1\over2} j}_{m\mu
M}Y^{(\ell)}_m (\theta,\phi)\ \chi_{\mu}$. Thus the spherical symmetry
reduces the number of amplitudes in the doublet even further from
four to three .

The Dirac eigenstates will then be an
eigenfunction of ${\vec L} \cdot {\vec L}$, $\vec J\cdot \vec J$, and $J_z$,

\bmath
\ba
{\vec J} \cdot {\vec J}\  \Psi_{n_r, \ell, j, M}^{s}({\vec
r}) =
j\ (j + 1)  \
\Psi_{n_r, \ell, j, M }^{s}({\vec r}),\\
{\vec L} \cdot {\vec L}\  \Psi_{n_r, \ell, j, M}^{s}({\vec
r}) =
\ell\ (\ell + 1)  \
\Psi_{n_r, \ell, j, M }^{s}({\vec r}),\\
J_z\  \Psi_{n_r, \ell, j,
M}^{s}({\vec r})= M  \  \Psi_{n_r, \ell, j, M }^{s}({\vec r}).
\label{LL}
\ea
\emath
The differential equation for $g_{n_r,\ell}(r)$ becomes:
\begin{equation}
\left [ {d^2 \over dx^2} - {\ell (\ell + 1) \over x^2} - x^2 + 
{{({\tilde E_N}^2 -
{\tilde M}^2)}\over \left [({\tilde E_N} + {\tilde M}) \ {{\tilde 
M}\omega^2 \over 2}
\right ]^{1\over 2}}\right ]\ r\  g_{n_r,\ell}({r}) = 0,
\end{equation}
where ${\tilde E_N} = E_N - V^0_V,\ {\tilde M} = M + V^0_S,\  x = 
\lambda\ r$, and
\bmath
\ba
\lambda
=
\left [({\tilde E_N} + {\tilde M}) \ {{\tilde M}\omega^2
}\right ]^{1\over 4}
\label {lambda}
\ea
\ba
{\tilde E_N^2} - {\tilde M}^2 = 2\ \lambda^2\ (N + {3\over2})
\ea
\emath
\subsubsection{Eigenfunctions}
The solutions to this
differential equation are well known and lead to the
upper amplitudes of the eigenfunctions
\begin{equation}
{ g}_{n_r,\ell}({r}) = {\cal{N}}(E_{N,\ell})\  \sqrt{{ 2\ \lambda^3\ 
n!\over \Gamma(\ell + n
+ {3\over2})}}\ e^{-{{x^2}
\over 2}} x^{\ell}L_{n_r}^{(\ell +{1\over 2})}(x^2),
\end{equation}
where ${\cal N}(E_{N})$ is the normalization determined by 
$\int_0^{2\pi} d\phi$ $
\int_{-\pi}^{\pi}sin(\theta) d\theta$ $ \int_0^{\infty} $ $ r^2\ dr 
(|{g}_{n_r,\ell}({\vec
r})|^2$ $ + \ |{f}_{n_r,\ell,j}({\vec r})|^2)$ = 1 and is the same 
function as given in (\ref {norm}). Clearly ${ g}_{n_r,\ell}({r})$
has $n_r$ nodes in the radial direction.

The lower components are
\begin{eqnarray}
&&{f}_{n_r,\ell,j = \ell -{1\over 2}}({\vec r}) =  \nonumber \\ &&-{{ {\cal N}(E_{N}})\over
{\tilde M} + {\tilde E_{N}}}\
\sqrt{{ 2\ \lambda^5\ n!\over \Gamma(\ell + n_r
+ {3\over2})}}\ e^{-{{x^2}
\over 2}} x^{\ell-1}( (n_{r} + 1)\ L^{(\ell - {1\over2})}_{n_{r}+ 
1}(x^2)+ (n_{r} + \ell + {1\over2})\
L^{(\ell - {1\over2})}_{n_{r}}(x^2)),
\end{eqnarray}
\begin{equation}
f_{n_r,\ell,j = \ell +{1\over 2}}({\vec r}) = \\ {{ {\cal N}(E_{N}})\over
{\tilde M} + {\tilde E_{N}}}\
\sqrt{{ 2\ \lambda^5\ n!\over \Gamma(\ell + n_r
+ {3\over2})}}\ e^{-{{x^2}
\over 2}} x^{\ell+1}( L^{(\ell + {3\over2})}_{n_{r} }(x^2)+  L^{(\ell 
+ {3\over2})}_{n_{r} -
1}(x^2))
\label {effas}
\end{equation}
Clearly the function ${f}_{n_r,\ell,j = \ell -{1\over 2}}({\vec r})$ 
has $n_r + 1$ nodes, one more node
than the upper component. The amplitude
${f}_{n_r,\ell,j = \ell +{1\over 2}}({\vec r}))$ has the same number of nodes
as the upper component. This agrees with the general theorem relating 
the number of radial nodes of the
lower comonents to the number of radial nodes of the upper 
component\cite{ami2}.

\subsubsection{Energy Eigenvalues}

The eigenvalue equation is

\begin{equation}
\sqrt{{\cal E}_N + 1}\ ({\cal E}_N - 1) = C\ (N + {3 \over 2}),
\end{equation}
where ${\cal E}_N = {{\tilde E}_N \over {\tilde M}}, C ={ {2}\ 
\omega \over
{\tilde M}}$, and
$N$ is the total oscillator quantum number, $N = 2n + \ell = 0, 1,\dots$.
We note that there is not only a degeneracy due to spin symmetry but 
there is also the
usual degeneracy of the non-relativistic harmonic oscillator; namely, 
that the energy
depends only on the total harmonic oscillator quantum number and the 
states with orbital angular momentum $\ell =
N, N-2, \dots, 0$ or 1 and angular momentum projection $m = \ell, \ell -1, \dots, \ell$ are all
degenerate.

Again the eigenvalue is:
\begin{equation}
E_N =  {\tilde M} \left [ B(A_{N}) + {1\over 3} + {4\over 9\ 
B(A_{N})} \right] + V_V^0,
\label {E1sph}
\end{equation}
and $A_{N} = C(N + {3 \over 2})$.
\subsection{Energy Spectrum}

\begin{figure}

\hspace*{-9mm}

\includegraphics[width=10cm]{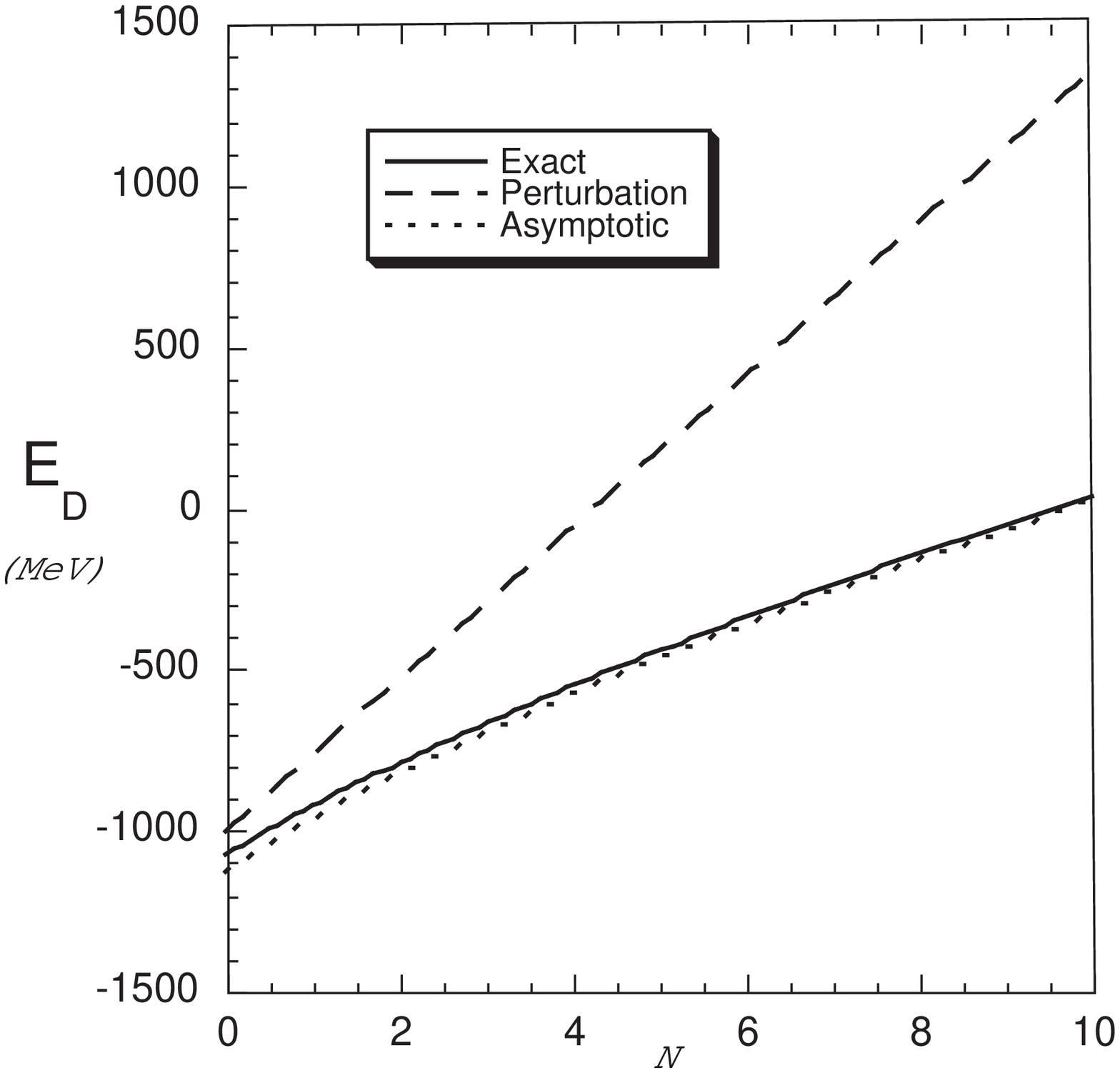}


\caption{
The Dirac binding energies, $E_D$, for the spherical harmonic oscillator as a function of $N$. 
The exact enegies are the solid line, the perturbation appoximation is the dashed line, and the
asymptotic approximation is the short dashed line. }\label{ho}
\end{figure}

In Figure \ref {ho} we plot the spherical harmonic oscillator Dirac binding enegies
$E_D = E_N- M$ with $E_N$ given in Eq. (\ref {E1sph}), the solid curve, as a function of
$N$. We chose the parameters to fit the lowest eigenenergies of the spectrum of an anti-proton
outside of $^{16} O$ in the relativistic mean field aprroximation \cite {thomas} and they are $C
=1.33, {\tilde M} = 252 $ MeV, and $V^0_V = -677 $ MeV. 

The dashed curve is $E_D$ using the pertubation approximation of $E_N$ given in Eq. (\ref
{pert}). The short-dashed curve is $E_D$ using the asymptotic limit of $E_N$ given in Eq.
(\ref {asymp}). Clearly the eigenenergies are in the relativistic asymptotic regime and not the
linear regime of the non-relativistic harmonic oscillator.

\begin{figure}


\includegraphics[width=8cm]{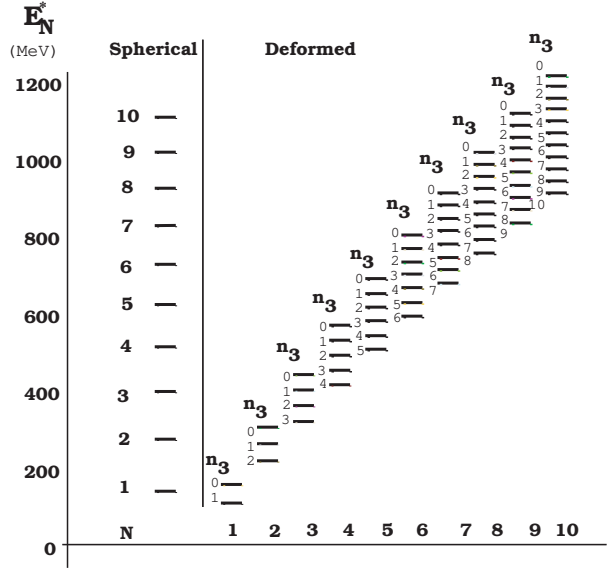}

\hspace*{9mm}
\caption{
On the left the excitation energies, $E_N^*$, for the spherical harmonic oscillator are plotted for
each
$N$. Each level  has a $(N+2)\ (N+1) $ degeneracy because of spin symmetry and because the allowed
orbital angular momenta are
$\ell = N, N-2,
\dots 0$ or $1$ and the allowd orbital angular momentum projections are $m = \ell, \ell - 1,
\dots -\ell$. On the right the excitation energies, $E^*$, for the deformed harmonic oscillator are
plotted in staggered groups for each $N$. Each group contains the levels for  $n_3 = 0,1,\dots,
N$. Each level has a $2\ (N -n_3) + 1$ degeneracy
for $(N -n_3)$ even  and a 
$2\ (N -n_3 + 1)$ degeneracy for $(N -n_3)$ odd because of spin symmetry and because the allowed
orbital angular momentum projections are
$\Lambda =
\pm (N -n_3),\pm (N -n_3 -2),
\dots \pm1$ or $0$.}\label{spectrum}
\end{figure}

In Figure \ref {spectrum} we plot the spherical harmonic oscillator excitation energies  $
{E}^*_N = E_N - E_0 $ for different $N$ on the far left. Each level  has a $(N+2)\ (N+1) $
degeneracy because of spin symmetry and because the allowed orbital angular momenta are
$\ell = N, N-2,
\dots 0$ or $1$ and the allowed orbital angular momentum projections are $m = \ell, \ell - 1,
\dots -\ell$. In the right of  Figure \ref {spectrum}
we plot the deformed excitation energies $
{ E}^*_{N, n_3} = E_{N,n_3} - E_{0,0}$. The deformed excitation energies are
staggered in groups for each $N$ and each group contains the levels for  $n_3 = 0,1,\dots, N$ with
the excitation energy increasing with decreasing $n_3$. The dimensionless oscillator strengths
are determined by $C^3 = C_{\perp}^2\ C_3$ and assuming a deformation $\delta = 0.33$ which leads
to $C_{\perp}= 1.49, C_3 = 1.05$ \cite {bohr}.  Each level has a $2\ (N -n_3) + 1$ degeneracy
for $(N -n_3)$ even  and a 
$2\ (N -n_3 + 1)$ degeneracy for $(N -n_3)$ odd because of spin symmetry and because the allowed
orbital angular momentum projections are
$\Lambda =
\pm (N -n_3),\pm (N -n_3 -2),
\dots \pm1$ or $0$. The splitting of the levels within each $N$ appears to be approximately linear
with $n_3$.

\subsection{Relativistic Contribution}
The normalization ${\cal N}(E)$ has the same functional form 
independent on whether the harmonic
oscillator is triaxial, axially deformed, or spherical. This 
normalization has also
been calculated independently by using
$\int_{-\infty}^{\infty}dx
\int_{-\infty}^{\infty}dy\int_{-\infty}^{\infty}dz\ 
(|{f}_{\mu}^+({\vec r})|^2 + |{f}_{\mu}^-({\vec r})|^2)
=
\int_{-\infty}^{\infty}dx
\int_{-\infty}^{\infty}dy\int_{-\infty}^{\infty}dz\ {g}({\vec
r})^*\ {p^2\over ({\tilde E} + {\tilde M})^2}\ {g}({\vec r})$ and we 
find agreement between the two different ways of calculating ${\cal N}(E)$.

This also tells us
that the probability of the lower component to the upper component is given by:
\begin{equation}
R_k = {\int_{-\infty}^{\infty}dx
\int_{-\infty}^{\infty}dy\int_{-\infty}^{\infty}dz\ 
(|{f}_{k,\mu}^+({\vec r})|^2 + |{f}_{k,\mu}^-({\vec r})|^2)
\over \int_{-\infty}^{\infty}dx
\int_{-\infty}^{\infty}dy\int_{-\infty}^{\infty}dz\ |{g}_k({\vec 
r})|^2}= {{\tilde E_k} - {\tilde M} \over
2\ ({\tilde E_k} + {\tilde M}) }.
\end{equation}
Thus for ${\tilde E}_k \approx {\tilde M}$ the system is not very 
relativistic and the contribution of the
lower components is small. For ${\tilde E}_k >> {\tilde M}$, this ratio 
approaches $1\over 2$. For free
particles this ratio approaches unity which indicates that the 
harmonic oscillator reduces the
relativistic effect.

\begin{figure}

\hspace*{-9mm}

\includegraphics[width=10cm]{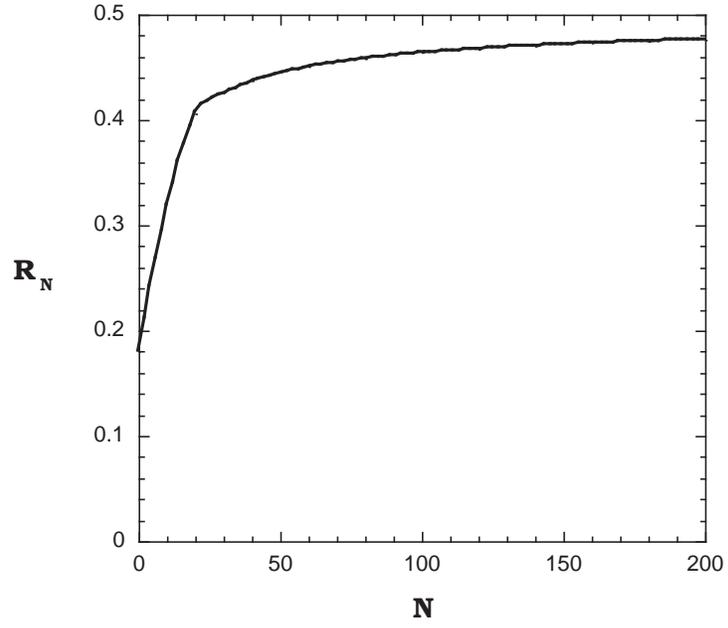}


\caption{
The ratio of the probability  of the lower components to the upper components, $R_N$, for the
spherical harmonic oscillator as a function of
$N$.  }\label{RN}
\end{figure}

In Figure \ref {RN} we plot this ratio for the spherical harmonic oscillator, $R_N $, as a 
function of $N$. Even for the most bound states this probability is about 20 \% and thus the
antinucleon  bound inside the nucleus is much more relativistic than a nucleon inside a nucleus for
which  this probability is about 1 \%.

\section{Higher Order Symmetry}

The non-relativistic spherical harmonic oscillator has an $SU(3)$ 
symmetry \cite {phil}. This symmetry is generated by the orbital angular momentum
operators ${\vec \ell}$ and the quadrupole operators

\begin{equation}
{ q}_m = 
{[rr]_m^{(2)}}  \lambda_{NR}^2 + {[pp]_m^{(2)} \over \lambda_{NR}^2} ,
\label {NR}
\end{equation}
where ${[rr]_m^{(2)}}$ means coupled to angular momentum rank 2 and projection $m$ and
$\lambda_{NR} = \sqrt {M \omega}$. This quadrupole operator is then a function of the dimensionless
variable 
$\vec x_{NR} = 
\lambda_{NR}\ \vec r$. These generators connect the degenerate states of the harmonic 
oscillator.

The same degeneracy that appears in the non-relativistic
spectrum appears in the relativistic  spectrum.
The upper component of the relativistic eigenfunction given in Eq. (\ref {ssph}) has the same
form as the non-relativistic harmonic oscillator eigenfunction except that it is a function of the
relativistic dimensionless variable $\vec x = 
\lambda  \vec r$. Therefore if the generators are written in terms of the  relativistic
dimesionless variable they will connect the upper components of all the degenerate states of
the relativistic harmonic oscillator in the same manner as the non-relavitistic quadrupole
operator in Eq. (\ref {NR}) connects the degenerate eigenstates of the non-relativistic harmonic
oscillator. Likewise, since the lower components are proportional to 
$U_p$ operating on the upper components, the dimensionless quadrupole
operator transformed by 
$U_p$ connects the lower components of the degenerate states in the same manner as the
non-relavitistic quadrupole operator in Eq. (\ref {NR}) connects the degenerate eigenstates of the
non-relativistic harmonic oscillator. However $\lambda$ depends on the energy ( see Eq. (\ref
{lambda})) and therefore the relativistic quadrupole generator is 
\begin{equation}
{ Q_m} = \left( \begin{array}{cc}
{[rr]_m^{(2)}} & 0 \\
0 & { U_p\ [rr]_m^{(2)}\ U_p}  \end{array} \right)
 \omega \sqrt{{\tilde M} (H_s - V_V^0 + {\tilde M}) } +  [pp]_m^{(2)}  {1\over \omega} \sqrt{{\tilde M} (H_s - V_V^0 +  {\tilde M})}\end{equation} since $U_p$ commutes with $\vec p$.
These generators
along with the orbitial angular momentum,
${\vec L}$,  given by (\ref {L}), connect the degenerate states with each other.
Work on the algebra is in progress.  For $M \rightarrow 
\infty, Q_m \rightarrow {\bar Q}_m$,
\begin{equation}
{\bar Q}_m = \left( \begin{array}{cc}
{[rr]_m^{(2)}} & 0 \\
0 & { U_p\ [rr]_m^{(2)}\ U_p}  \end{array}
    \right)\ {\omega\ { M} } + 
{[pp]_m^{(2)}\over \omega\
{ M}}\ ,
\end{equation}
which forms an $SU(3)$ algebra with ${\vec L}$.
\section{Summary and Conclusions}
We have derived the eigenfunctions and eigenenergies for a Dirac 
Hamiltonian with
triaxial, axially deformed, and spherical harmonic oscillator 
potentials and with equal scalar and vector
potentials. In all cases the Dirac Hamiltonian is invariant with 
respect to the $SU(2) $ spin symmetry and
thus the eigenenergies are independent of spin. For axially symmetric 
potentials the Dirac Hamiltonian is
invariant with respect to a $SU(2) \times U(1)$ group and the 
eigenenergies are degenerate with respect
to the orbital angular momentum projection along the
$z$-axis which generates the $U(1)$. For the spherical oscillator the 
eigenenergies are degenerate with
respect to the orbital angular mometum and hence invariant under the $SU_L(2)
\times SU(2)$ group. These energies also have a higher degeneracy which is the same as the
non-relativistic harmonic oscillator; that is, they depend only on the total harmonic 
oscillator quantum number. The 
generators that connect these degenerate states have been derived but a larger symmetry group
analogous to
$SU(3)$ symmetry has yet to be identified.
However, for infinite mass, the spherical relativistic harmonic 
oscillator is invariant with respect to an
$SU(3) \times SU(2)$ group.

The eigenenergies have the same functional form for triaxial, axially deformed, and
spherical potentials and depend on one variable which is a linear combination of the 
oscillator quanta in a given
direction weighted by the strength of the oscillator potential in 
that direction. The spectrum is infinite and the eigenenergies and are linear in the
harmonic oscillator quanta for small oscillator strength but 
increase slower than linear for large oscillator strength. 

\section{Acknowledgements}
The author would like to thank T. B\"urvenich, R. Lisboa, M. Malheiro, A. S. de Castro, P.
Alberto and M. Fiolhais for discussions. This work  was supported by the U.S. Department of Energy
under contract W-7405-ENG-36.
\pagebreak

\pagebreak


\end{document}